\documentstyle[11pt,epsfig,epsf,axodraw]{article}
\newcommand{\ba}{\begin{array}}
\newcommand{\ea}{\end{array}}
\newcommand{\bd}{\begin{displaymath}}
\newcommand{\ed}{\end{displaymath}}
\newcommand{\be}{\begin{equation}}
\newcommand{\ee}{\end{equation}}
\newcommand{\bea}{\begin{eqnarray}}
\newcommand{\eea}{\end{eqnarray}}

\def\bra{\langle}
\def\ket{\rangle}

\def\e{\epsilon}

\def\q2 {q^2}

\begin{document}
\begin{flushright}
{\large MRI-P-001107 \\
November, 2000}  
\\ hep-ph/0011375
\end{flushright}
\vskip1cm
\begin{center}
{\Large\bf  Signals of R-parity violating supersymmetry in neutrino scattering at 
muon storage rings}\\[15mm]
Anindya Datta\footnote{E-mail: anindya@mri.ernet.in}, 
Raj Gandhi\footnote{E-mail: raj@mri.ernet.in}, 
Biswarup Mukhopadhyaya\footnote{E-mail: biswarup@mri.ernet.in}\\
{\em Harish-Chandra Research Institute (formerly, Mehta Research Institute),
\\
Chhatnag Road, Jhusi, Allahabad - 211 019, India}, 
\\[7mm]
Poonam Mehta\footnote{E-mail: mpoonam@mri.ernet.in}\\
{\em Department of Physics and Astrophysics, University of Delhi, Delhi - 110007} 
\end{center}

\vskip 17pt
\begin{abstract}
  
  Neutrino oscillation signals at muon storage rings can be faked by
  supersymmetric (SUSY) interactions in an R-parity violating scenario. We
  investigate the $\tau$-appearance signals for both long-baseline and
  near-site experiments, and conclude that the latter is of great use in
  distinguishing between oscillation and SUSY effects.  On the other hand, for
  a wide and phenomenologically consistent choice of parameters, SUSY can
  cause a manifold increase in the event rate for wrong-sign muons at a
  long-baseline setting, thereby providing us with signatures of new physics.

\end{abstract}

\vskip 1 true cm

\setcounter{footnote}{0}

The increasingly strong empirical indications of neutrino oscillations from
the observed solar and atmospheric neutrino deficits \cite{superk_results}
have emphasised the need for their independent confirmation in accelerator and
reactor experiments. One of the actively discussed possibilities in this
connection is a muon storage ring \cite{storage_ring,albright} which can act
as an intense source of collimated neutrinos impinging upon a fixed target. A
$\mu^-$ ($\mu^+$) beam can thus produce both $\nu_\mu$ (${\bar \nu_\mu}$) and
${\bar \nu_e}$ ($\nu_e$), thereby providing an opportunity to test both
$\nu_e$-$\nu_\mu$ and $\nu_\mu$-$\nu_\tau$ oscillations which are the favoured
solutions for the two anomalies mentioned above.

In the simplest extensions of the standard model, non-degenerate
masses for the different neutrino species (and consequent mixing among
them) can account for the oscillation phenomena. Considering, for
example, the atmospheric $\nu_\mu$ deficit, the SuperKamiokande (SK)
results \cite{superk_results} strongly suggest $\nu_\mu$-$\nu_\tau$
oscillation with $\Delta m^2~\simeq~10^{-3}-10^{-2}~eV^2$ and $\sin^2
2\theta~\simeq~1$.  Such oscillation was earlier indicated by the
Irving-Michigan-Brookhaven and Kamiokande collaborations, and has been
also supported more recently by the SOUDAN-II \cite{soudan_results}
and MACRO \cite{macro_results} experiments.  At a muon storage ring,
one therefore expects a certain fraction of the $\nu_\mu$'s to
oscillate into $\nu_\tau$, depending on the energy and the baseline
length.  Interaction of these $\nu_\tau$'s with the target material
will produce $\tau$-leptons, the detection of which may, in the
simplest case, be interpreted as additional proof of oscillation
\cite{tau_appearence}. Similarly, the detection of `wrong sign muons'
may be a vindication of the ${\bar\nu_e}$-${\bar \nu_\mu}$ oscillation
hypothesis, thereby providing one with a probe of the parameter spaces
corresponding to the vacuum and matter-enhanced oscillation solutions
to the solar neutrino puzzle.

However, the predicted rates of $\tau$-appearance or wrong-sign muons
in a given experimental setting can be significantly affected by
non-standard interactions. In other words, it is possible for
non-oscillation physics to intervene and fake oscillation phenomena.
Lepton flavour violation effects in general can mimic the neutrino
oscillation signal. Such issues have been already discussed in some
detail in a model-independent way in the refs. \cite{lepton_flavour}.
For example, it is possible for {\em un-oscillated $\nu_\mu$'s} to
scatter into $\tau$'s in an R-parity violating supersymmetric (SUSY)
framework (with $R~=~(-1)^{(3B + L + 2S)}$), by virtue of
lepton-number violating trilinear couplings \cite{r_parity}.  Also,
such couplings can produce ${\bar \nu_\mu}$'s from $\mu^{-}$-decay and
thus give rise to $\mu^{+}$'s in the detector even in the absence of
oscillation \cite{r_par_neu}.  It is important to know the effects of
such interactions for two reasons: {\it (i)} to look for enhancement
in $\tau$ and wrong-sign muon event rates and thus to uncover SUSY
effects, and {\it (ii)} to see to what extent the signals supposedly
coming from oscillation are faked by such new physics. In this paper,
we show that one can answer both questions by combining long-baseline
experiments with those in which one places the neutrino detectors at a
short distance from the storage ring, where the oscillation
probability gets suppressed by the baseline length.

In addition, other non-standard options such as left-right symmetric models
and theories with extra gauge bosons also can lead to some of the observable
consequences discussed here. However, the rather stringent lower bounds (of
the order of 500 $GeV$ and above) on the masses of these bosons suppress the
contributions. On the other hand, the very relaxation of lepton number
conservation in a SUSY scenario introduces several additional couplings in the
theory, not all of which can be excluded with a great degree of severity from
currently available experimental results. It is some of these new interaction
terms, coupled with the possibility of having sfermions in the mass range of
100-300 $GeV$, that are responsible for the remarkable enhancement of tau-and
wrong-sign muon event rates at a neutrino factory. Similarly, lepton number
violating couplings also occur in the theories with leptoquarks. In R-parity
violating SUSY (with $\lambda '$-type couplings), squarks behave in much the
same way as scalar leptoquarks. Therefore, the results of our analysis on
$\tau$ appearance are equally applicable to the interactions involving scalar
leptoquarks with charge $\pm \frac{1}{3}$.

While a long-baseline experiment is of advantage from the viewpoint of
oscillation, a detector placed at a near site has also been considered in
recent studies \cite{short_base,albright}. Its merit lies in having a large
event rate even when its dimensions are small.  In reference \cite{albright},
for example, $\tau$ detection has been investigated in the context of a
detector consisting of an array of tungsten sheets with silicon tracking. Such
a detector will be helpful in isolating new $\tau$-producing interactions
which can be potential contaminants of the oscillation signature.

Let us first consider $\tau$-appearance through oscillation.  For a
muon-neutrino with energy $E_\nu$ (in $GeV$) traversing a distance $L$
(in $km$), the probability of oscillation into a tau-neutrino is given
by

\begin{equation}
{\cal P}_{\nu_{\mu,e}\rightarrow \nu_\tau} = \sin^2 2\theta\,\, \sin^2\left[ 
1.27\, \Delta m^2\, {L\over E_\nu}\right]
\end{equation}

\noindent
where $\Delta m^2$ is the mass-squared difference between the
corresponding physical states in $eV^2$, and $\theta$, the mixing
angle between flavours.  For a baseline length of, say, 700 $km$, and a
muon beam of energy 50 $GeV$, this probability corresponding to the
solution space for the atmospheric SK data lies in the range
$10^{-3} - 10^{-2}$, and is smaller for shorter baselines.  In order
to obtain the $\tau$-event rate one has to fold the charged current
cross-section with this probability as well as the $\nu_\mu$ energy
distribution, and finally use the effective luminosity appropriate for
the cone subtended by the detector, which depends on the area and baseline
length.

The presence of scalars carrying lepton ($L$) or baryon ($B$) number
in a SUSY scenario makes it possible to have {\em one} of the above
quantum numbers broken while the other is conserved. One can thus
avoid undesirable consequences like fast proton decay, and can still
be consistent with all other symmetries when R-parity is violated.  In
such a scenario with broken lepton number, the corresponding part of
the superpotential is given (suppressing colour and $SU(2)$ indices)
by \cite{r_par_neu},

\begin{equation}
W_{\not L} =  \epsilon_i {\hat L}_i {\hat H}_2 +            
\lambda_{ijk} {\hat L}_i {\hat L}_j {\hat E}_k^c +
\lambda_{ijk}' {\hat L}_i {\hat Q}_j {\hat D}_k^c 
\end{equation}  

\noindent
where i, j and k are generation indices, $\hat L$ and $\hat Q$ represent the
SU(2) doublet lepton and quark superfields, and $\hat E$, $\hat U$ and $\hat
D$ denote the right-handed charged lepton, up-type quark and down-type quark
superfields respectively. In terms of the component fields (with the sfermion
fields characterised by the tilde sign), the trilinear terms above lead to
interactions of the form

\bea
{\cal {L}} &=&  \lambda'_{ijk} ~\big[ ~\tilde d^j _L \,\bar d ^k _R \nu^i _L
  + (\tilde d ^k_R)^\ast ( \bar \nu ^i_L)^c d^j _L +
   \tilde \nu ^i _L \bar d^k _R d ^j _L  \nonumber \\
& & ~~~~~~~ -\tilde e^i _L \bar d ^k _R u^j _L
- \tilde u^j _L \,\bar d ^k _R e^i _L 
-(\tilde d^k _R)^\ast (\bar e ^i _L)^c u^j _L \big] + h.c. \nonumber \\
& & + \lambda_{ijk} ~\big[ ~\tilde e^j _L \,\bar e ^k _R \nu^i _L
  + (\tilde e ^k_R)^\ast (\bar \nu ^i _L)^c e^j _L +
   \tilde \nu ^i _L \bar e^k _R e ^j _L  - (i \leftrightarrow j) \big] + h.c
\eea

It should be noted that these interaction terms violate both lepton
flavour and lepton number. By suitable combinations of two such terms,
it is possible to achieve contributions to processes which conserve
lepton number but involve transition between different
generations.  The implications of such interactions have been
investigated earlier in the contexts of solar \cite{sol} and
ultra-high energy neutrinos \cite{ultra}.  At a neutrino factory, they
can affect the $\tau$ or wrong sign $\mu$ event rates in the 
following ways:

\begin{enumerate}
\item $\lambda'$-type interactions give rise to a $\tau$ starting from
a $\nu_\mu$ which is produced via standard muon decay.

\item $\lambda$-type interactions produce a ${{\bar \nu_\mu}}$ from muon decay,
which subsequently has standard charged-current interaction with the target, 
leading to a wrong-sign muon.

\item A $\nu_\tau$ can be produced as a result of $\lambda$-type interactions
in muon decay, which produces a $\tau$ through standard interaction.

\item The ${\bar \nu_e}$ from $\mu^-$ decay may scatter into a $\mu^+$ via
$\lambda'$-type interactions with the target. 
\end{enumerate}

Here we present results for cases 1 and 2 above. The Feynman diagrams
for these two cases are shown in figure 1, where we have chosen those
$\lambda'$-and $\lambda$-interactions which make $\tilde{b}$ and
$\tilde{\tau}$ the mediators in the corresponding diagrams.
Predictions for cases 3 and 4 are qualitatively similar to those for 1
and 2 respectively.

\begin{figure}
\begin{picture}(280,100)(0,0)
\vspace*{- 1in}
\ArrowLine(60,-5)(100,30)
\Text(60,-16)[r]{$\nu_\mu$}
\ArrowLine(57,65)(100,30)
\Text(60,80)[r]{$d$}
\DashLine(100,30)(170,30){4}
\Text(135,40)[b]{${\tilde b}_R$}
\ArrowLine(170,30)(200,70)
\Text(202,80)[b]{$u$}
\ArrowLine(170,30)(210,-3)
\Text(202,-16)[b]{$\tau$}
\Text(200,30)[b]{$\lambda '_{313}$}
\Text(70,30)[b]{$\lambda '_{213}$}
\Text(135,-15)[b]{(a)}

\ArrowLine(245,30)(305,30)
\Text(260,21)[r]{$\mu^-$}
\ArrowLine(305,30)(335,75)
\Text(325,70)[r]{$\nu_e$}
\DashLine(305,30)(370,30){4}
\Text(345,40)[b]{${\tilde \tau}_L$}
\ArrowLine(370,30)(400,70)
\Text(402,76)[b]{$e^-$}
\ArrowLine(370,30)(410,-3)
\Text(402,-10)[b]{${\bar\nu}_\mu$}
\Text(400,30)[b]{$\lambda _{231}$}
\Text(288,40)[b]{$\lambda _{132}$}
\Text(335,-15)[b]{(b)}
\end{picture}
\vskip .3in
\caption{ {\em Feynman diagrams for  processes producing (a) a $\tau$
or (b) a wrong sign muon in R-parity violating SUSY.}
}
\end{figure}
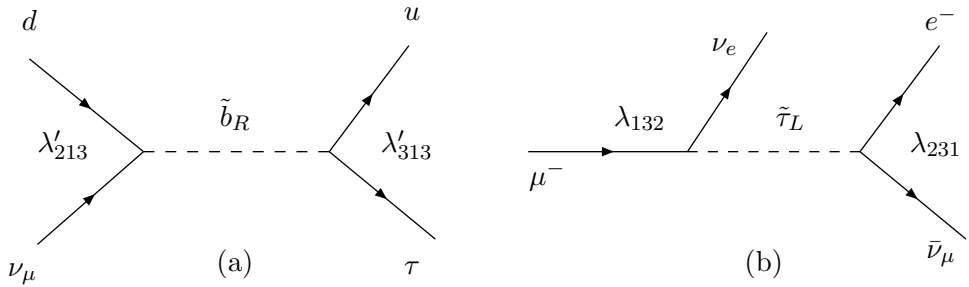

The standard model charged current cross-section for $\nu_\tau N
\longrightarrow \tau^- X$ can be found, for example, in references
\cite{albright, tau_appearence}. In R-parity violating SUSY, a
$\nu_\mu$ can give rise to the same final state through the couplings
$\lambda'_{213}$ and $\lambda'_{313}$ when the tree-level process is
mediated by a b-squark. From considerations of phase-space
availability as well as parton densities in a nucleon, the most
favourable $\tau$-producing processes at the quark level are $\nu_\mu
d \longrightarrow \tau^{-} u$ and $\nu_\mu {\bar{u}} \longrightarrow
\tau^{-} {\bar{d}}$.  On performing Fierz transformations on the SUSY
amplitudes for these processes, one obtains

\bea
{\cal M}_{SUSY}(\nu_\mu \;d \longrightarrow \tau^- \; u) &=& 
\frac{\lambda'_{213}\lambda'_{313}}{2 (\hat s - m ^2_{\tilde b_R})}\;
\big[\bar u_{\tau} \gamma_{\mu} P_L u_{\nu_\mu}\big]\,
\big[\bar u_{u} \gamma^{\mu} P_L u_d \big] \nonumber \\
{\cal M}_{SUSY}(\nu_\mu \;\bar u \longrightarrow \tau^- \; \bar d) &=& 
\frac{\lambda'_{213}\lambda'_{313}}{2 (\hat t - m ^2_{\tilde b_R})}\;
\big[\bar u_{\tau} \gamma_{\mu} P_L u_{\nu_\mu}\big]\,
\big[\bar v_{u} \gamma^{\mu} P_L v_d \big] 
\eea
\noindent
where $m_{\tilde{b}}$ is the b-squark mass.
Left-right mixing in the squark sector has been neglected here.
No consequence of the phases of the $\lambda'$-type couplings has 
been considered.
It should be noted that the ($\nu_\mu \leftrightarrow \nu_\tau$)
oscillation amplitude (arising from neutrino mass splitting) is 
purely imaginary in a two
level analysis \cite{palash}. Hence, there is no interference between
the oscillation and R-parity violating amplitudes as long as the
product of two $\lambda'$ couplings is real. We have worked under
such an assumption here.

There are phenomenological bounds on the $L$-violating couplings
\cite{bounds_r_par}; however, most of these bounds are derived on the
assumption that only a single coupling at a time is non-zero (we call
them the `stand-alone' bounds). Thus, although there are individual
limits on $\lambda'_{213}$ and $\lambda'_{313}$, obtained from the
universality of charged-current decays of $\pi^-$ and $\tau$
\cite{bounds_r_par}, such limits are not necessarily applicable in the
most general case.  At any rate, no conclusive limit has been obtained
for the product ($\lambda'_{213}$ $\lambda'_{313}$). Hence this
product can be treated as a {\em free parameter when it comes to
  looking for experimental signals}.  We have also checked that the
values of the {\it effective lepton flavour-violating coupling} used
here do not contradict any limits on such couplings available in the
literature.

\begin{figure}[ht]
\centerline{ 
\epsfxsize=10cm\epsfysize=7.0cm
                     \epsfbox{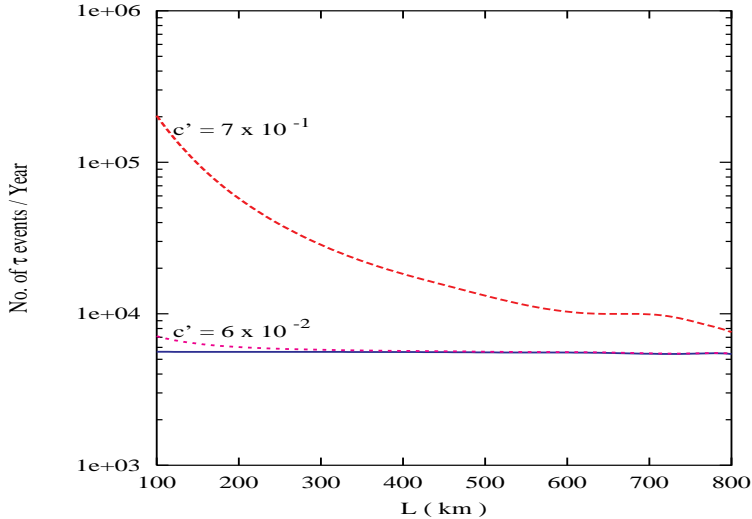}
}

\caption{{\em The $\tau$-event rate as a function of the baseline length
(L) for $E_\mu = 50 ~GeV$. The solid line corresponds to the contribution from 
$\nu_\mu-\nu_\tau$ oscillation, 
using SK parameters (see text). The two dashed lines refer to cases where
the SUSY contributions are included, taking different values of 
$c' ~(\equiv \lambda'_{213} \lambda'_{313})$.  
 }}
\label{fig:fig1}
\end{figure}

In figure \ref{fig:fig1} we show the event rates for tau production
for long-baseline experiments, as functions of the baseline length,
for a 50 GeV muon beam. When calculating the total $\tau$ appearance
rate, we add the R-parity violating contribution with the standard model
contribution (from $\nu_\tau ~N \rightarrow \tau^- ~X$) after folding
the latter with oscillation probability. The results presented here
correspond to a sample detector of mass $10~kT$, with a circular
cross-section of $100~m^2$. Such a specification is similar to that of
the ICANOE experiment \cite{icanoe}.  A muon source producing
$10^{20}$ muons per year has been assumed.  The expected event rates
for both standard charged current and SUSY contributions added to them
are displayed.  We have used CTEQ4LQ \cite{cteq} parton distributions
to calculate the event rates.  The standard model contribution to the
$\tau$-production rates has been calculated assuming an oscillation
probability corresponding to $\Delta m_{23}^2~\simeq~5.0 \times
10^{-3}~eV^2=$ and $\sin^2 \theta_{23} ~=~1$, which is within
 solution space for the atmospheric $\nu_\mu$
deficit. An average $\tau$-detection efficiency of 30\%
\cite{albright,icanoe} has been used here.  Different values of the
products of the R-violating couplings have been used, with a
bottom-squark mass of $300~GeV$, which is consistent with current
experimental limits.  The lower one of these corresponds to the
product of the stand-alone bounds of the individual couplings; we also
display the results with a value of $c'$ which is 10 times greater, and
is close to the product of the perturbative limits of the individual
couplings.  As can be seen from the figure, for baselines of length
$\ge~ 200~km$, R-parity violating effects make a serious difference
only when the couplings are close to their perturbative limits, while
for shorter baselines ($\simeq 100~km$), they can be competitive even
with values on the order of the stand-alone bounds.

However, new physics effects are quite clearly separated when one
comes to a near-site detector setting. Here the standard model
contribution is suppressed due to the paucity of $\tau$'s produced in
oscillation.  In figure 3 we show some plots of $\tau$-event rates
with a $1~kT$, $2500~cm^2$ detector placed at a distance of $40~m$
from the storage ring \cite{albright}. The two sets of values for the
product of the $\lambda'$-type couplings already used in the previous
figure are also used here; in addition, we show the predictions for
two considerably smaller values of this product.  One notices a 
substantial enhancement in the number of $\tau$-events
(calculated again with an assumed average detection efficiency of
30\%) to a level considerably higher than what the standard model
predicts. This is the case even when the relevant coupling strengths
are much smaller than the limits given in reference \cite{allanach}.
Combining figures 2 and 3, the conclusion, therefore, is that {\em
  even when the couplings are well within the bounds for the
  stand-alone situation}, near-site effects arising from them lead to
overwhelmingly large $\tau$-production, while for long-baseline
experiments, contamination of the oscillation signals through
R-violating interactions is appreciable when one goes beyond the
limits derived on the assumption that only one coupling is
non-vanishing at a time.

\begin{figure}[ht]
\centerline{ 
\epsfxsize=10cm\epsfysize=7.0cm
                     \epsfbox{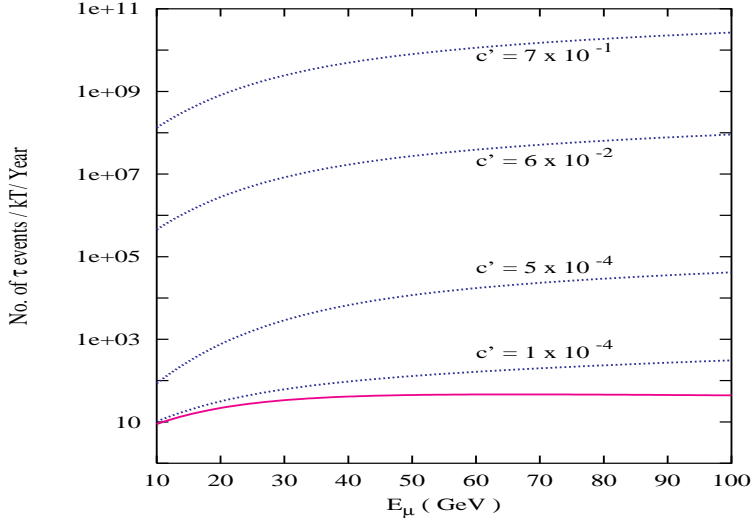}
}

\caption{\em{The $\tau$-event rate as function of the muon beam energy
for a near-site detector. The solid line shows the oscillation contribution,
while the SUSY contributions are included to the dashed lines, using 
different values of $c' ~(\equiv \lambda'_{213} \lambda'_{313})$.}}
\label{fig:fig2}
\end{figure}

Similarly, a $\nu_\tau$ can also be produced in the
decay of the $\mu^{-}$ via diagrams of the kind shown in figure 1. It
can consequently produce a $\tau$ at the detector even without oscillation. 
In such a case, the event rates are suppressed by the branching ratio of
the R-parity violating decay, to an extent
depending on the product of the corresponding $\lambda$-couplings. 
The predictions are similar to the ones discussed above.

\begin{figure}[ht]
\centerline{ 
\epsfxsize=10cm\epsfysize=7.0cm
                     \epsfbox{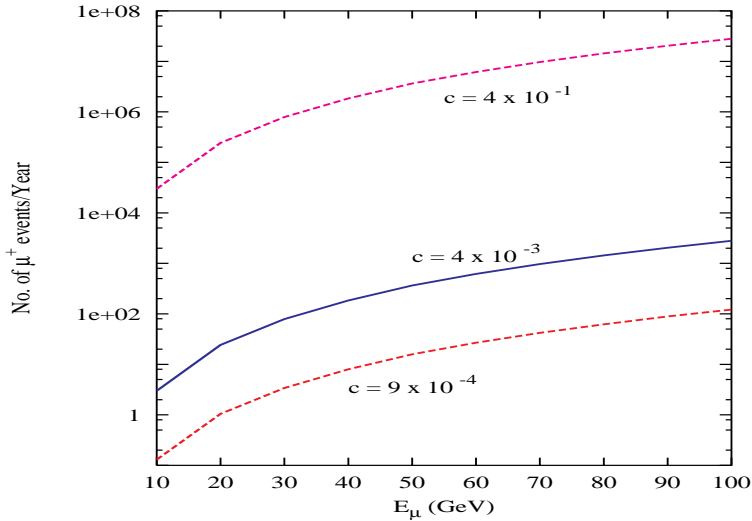}
}

\caption{{\it The event rates for wrong-sign muons as functions of 
muon energy, for a
baseline length of 250 km and a 10 kT detector of area 100 m$^2$. 
Different values of $c ~(\equiv \lambda_{231} \lambda_{132})$ have been
used, with $m_{\tilde \tau } = 100 ~GeV$}}
\label{fig:fig3}
\end{figure}

\begin{figure}[ht]
\centerline{ 
\epsfxsize=10cm\epsfysize=7.0cm
                     \epsfbox{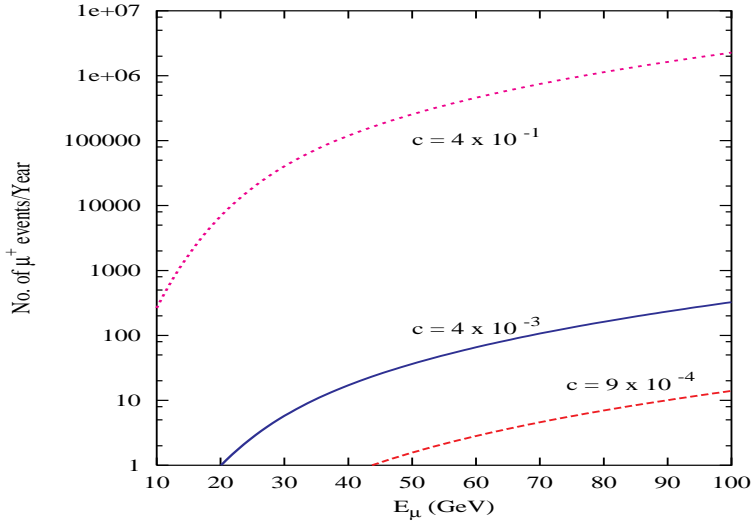}
}

\caption{{\it The event rates for wrong-sign muons as functions of muon 
energy, for a
baseline length of 732 km and a 10 kT detector of area 100 m$^2$. 
Different values of $c ~(\equiv \lambda_{231} \lambda_{132})$ have been
used, with $m_{\tilde \tau } = 100 ~GeV$}}
\label{fig:fig4}
\end{figure}

Next, we consider wrong-sign muons produced due to R-parity violating effects
in {\em muon decays}.  The Mikhyev-Smirnov-Wolfenstein (MSW) solution to the
solar neutrino problem with matter-enhanced $\nu_e$-$\nu_\mu$ oscillation
requires a mass-splitting of $\simeq~10^{-5}~eV^2$ between the mass
eigenstates \cite{solar}.  It has been found earlier \cite{tau_appearence}
that with a muon beam energy of upto $50~GeV$, and with standard charged
current interactions, one can hardly expect to see any events given this kind
of mass-splitting, for any realistic baseline length. The situation is even
worse for the vacuum oscillation solution which requires $\Delta
m^2~\simeq~10^{-10}~eV^2$. Thus a sizable event rate for wrong-sign muons at a
long-baseline experiment should be interpreted as a signal of some new effect,
unless $\nu_e$-$\nu_\mu$ oscillation is {\em not} the solution to the solar
neutrino puzzle. In the latter situation, however, the predicted wrong-sign
muon rates allow one to probe the solution space to, for example, the LSND
results \cite{lsnd}. In such a case, it becomes even more important to
understand the potential contributions coming from new physics effects such as
R-parity violating SUSY. In the discussion below, we have tried to demonstrate
our main point by confining ourselves to solar neutrino solution space, and
showing the visibility of the events through SUSY interactions.

Equation (3) tells us that the $\nu_\mu$ produced in $\mu$-decays
cannot give rise to a $\mu^+$ through R-parity-violating interactions
unless there is substantial left-right mixing in the (D-type) squark
sector.  On the other hand, diagrams of the kind shown in figure 1(b)
can lead to decays like $\mu^{-} \longrightarrow \nu_\e e {\bar
  \nu_\mu}$.  This decay is governed, for example, by the product
$\lambda_{231} \lambda_{132}$ when the process is mediated by a stau.
The decay amplitude, as obtained from figure 1, is

\be
{\cal M}_{SUSY}(\mu \longrightarrow \nu_e \;e \;{\bar {\nu_\mu}}) = 
\frac{\lambda_{132}\lambda_{231}}{(s_1 - m ^2_{\tilde \tau_L})}\;
\big[\bar u_{\nu_e} P_R u_{\mu}\big]\,
\big[\bar u_{e} P_L u_{\nu_\mu} \big]  
\ee
\noindent
where $s_1~=~(p_{\mu}~-~p_{\nu_e})^2$

In figures 4 and 5 we show the event rates for a typical ICANOE-type
detector as functions of the muon beam energy, using different values
of the above product. Predictions are made for two different baseline
lengths, one corresponding to the K2K proposal (figure 4) and the
other, to the Fermilab-SOUDAN or CERN-Gran Sasso long baseline
experiment.  Again, the value 0.004 corresponds to the product of the
stand-alone bounds.  In addition, two other values, one close to the
perturbative limit and the other one considerably smaller, have been
used. It may be noted that the only limit on the relevant product c,
attempted from the absence of muonium-antimuonium conversion
\cite{kim}, is about $6.3 \times 10 ^{-3}$.  Two of the three values
of c taken here are consistent with these limits.  Substantial event
rates are produced even with such values.  In addition, a general
limit on the parameter, $g^s_{RR}$
\footnote{The most general scalar interaction leading to the decay of a
muon to a right-handed electron can be written as: ${\cal L} =
\frac{4\,G_F}{\sqrt{2}}\,\left[g^S _{RR}\, \bra \bar e_R | \Gamma ^S
|(\nu_e) _L \ket\, \bra (\bar \nu_\mu)_L | \Gamma ^S |\mu_R\ket
\right]$. The current experimental limit on $g^S _{RR}$ constrains it
to be less than 0.066 \cite{pdg}.}  denoting the scalar coupling of a
muon leading to its decay into a right-handed electron exists in the
literature. Such a limit translates to $c < 0.022$, with which  
a large part of the parameter space covered
in the figures is consistent.

Even with conservative choices  of the interaction strengths, a 
clear prediction of ten to several hundred events can be
observed for $E_\mu~\simeq~50~GeV$, in the R-parity violating case,
while no events are expected so long as the masses and mixing in the $\nu_\mu
-\nu_\e$ sector offer a solution to the solar deficit.
 Clearly, a shorter baseline such as the one shown in
figure 4 is of greater advantage. Furthermore, a near-site detector should
be able to detect a huge abundance of events, although backgrounds caused
by muons produced upstream of the detector pose additional problems there. 
Taking everything into account, we conclude that that a far-site detector
with modest baseline-length like the one studied in figure 4 is probably the
optimal answer to questions on possible contributions to wrong-sign muon 
signals. This statement, however, ceases to be valid if the mass-squared 
splitting corresponding to $\nu_e$-$\nu_\mu$ oscillation belongs,
for example, to the solution space for the LSND results. A confirmation
(or otherwise) of the LSND claim is expected to come from the MiniBooNE 
experiment. In case of a reaffirmation of such kind of  $\nu_e$-$\nu_\mu$ 
oscillation, for which substantial wrong-sign muon rates are predicted in
long-baseline experiments, one has to worry about possible faking by SUSY
processes discussed here. Under such circumstances, one has to combine the
observed data with those obtained from a near-site detector to separate
the two types of effects.

In conclusion, we have investigated the effects of the R-parity
violating trilinear couplings on the suggested signals of neutrino
oscillations at a muon storage ring. We find that, while the new
couplings have to be on the higher side to show a detectable
enhancement in the $\tau$-appearance rate with long baselines, even
tiny R-violating couplings can lead to very large number of $\tau$'s
at a near-site detector, much in excess of what is expected via
oscillation. Near-site experiments can thus be recommended for
isolating new physics effects that fake signals of neutrino
oscillation.  On the other hand, a class of R-violating interactions,
with strengths well within their current experimental limits, can be
responsible for an enhanced rate of wrong sign muons at a
long-baseline experiment.  Since the solution space for the solar
neutrino puzzle does not permit such event rates, such muons, if
observed at a neutrino factory, can therefore be greeted as harbingers
of some new physics, of which R-parity violating SUSY is a 
favoured example.

\noindent
{\bf Acknowledgement:} We thank Sukanta Dutta for help at the initial
stage of this work.  P.M. acknowledges financial support from the
Council for Scientific and Industrial Research, India, and the
hospitality of Harish-Chandra Research Institute where this work has
been done.

\newcommand{\plb}[3]{{Phys. Lett.} {\bf B#1,} #2 (#3)}                  %
\newcommand{\prl}[3]{Phys. Rev. Lett. {\bf #1,} #2 (#3)}        %
\newcommand{\rmp}[3]{Rev. Mod.  Phys. {\bf #1,} #2 (#3)}             %
\newcommand{\prep}[3]{Phys. Rep. {\bf #1,} #2 (#3)}                     %
\newcommand{\rpp}[3]{Rep. Prog. Phys. {\bf #1,} #2 (#3)}             %
\newcommand{\prd}[3]{Phys. Rev. {\bf D#1,} #2 (#3)}       
\newcommand{\epjc}[3]{Euor. Phys. J. {\bf C#1,} #2 (#3)}                 %
\newcommand{\prc}[3]{{Phys. Rev.} {\bf C#1,} #2 (#3)}  
\newcommand{\np}[3]{Nucl. Phys. {\bf B#1,} #2 (#3)}                    %
\newcommand{\npbps}[3]{Nucl. Phys. B (Proc. Suppl.) 
           {\bf #1,} #2 (#3)}                                           %
\newcommand{\sci}[3]{Science {\bf #1,} (#3) #2}                 %
\newcommand{\zp}[3]{Z.~Phys. C{\bf#1,} #2 (#3)}                 %
\newcommand{\mpla}[3]{Mod. Phys. Lett. {\bf A#1,} #2 (#3)}             %
\newcommand{\ijmp}[3]{{ Int. J.  Mod. Phys.} {\bf A#1}, #2 (#3)}          %
 \newcommand{\apj}[3]{ Astrophys. J.\/ {\bf #1,} #2 (#3)}       %

\newcommand{\astropp}[3]{Astropart. Phys. {\bf #1,} #2 (#3)}            %
\newcommand{\ib}[3]{{ ibid.\/} {\bf #1,} #2 (#3)}                    %
\newcommand{\nat}[3]{Nature (London) {\bf #1,} (#3) #2}         %
 \newcommand{\app}[3]{{ Acta Phys. Polon.   B\/}{\bf #1,} (#3) #2}%
\newcommand{\nuovocim}[3]{Nuovo Cim. {\bf C#1,} (#3) #2}         %
\newcommand{\yadfiz}[4]{Yad. Fiz. {\bf #1,} (#3) #2;             %
Sov. J. Nucl.  Phys. {\bf #1,} #3 (#4)]}               %
\newcommand{\jetp}[6]{{Zh. Eksp. Teor. Fiz.\/} {\bf #1,} (#3) #2;
           {JETP } {\bf #4} (#6) #5}%
\newcommand{\philt}[3]{Phil. Trans. Roy. Soc. London A {\bf #1} #2  
        (#3)}                                                          %
\newcommand{\hepph}[1]{(hep--ph/#1)}           %
\newcommand{\hepex}[1]{ (hep--ex/#1)}           %
\newcommand{\astro}[1]{(astro--ph/#1)}         %

\end{document}